\documentclass{appolb}
\usepackage{graphicx}
\usepackage{subfigure}
\usepackage{url}
\usepackage{lineno}

\begin{document}
\title{Status of the Compressed Baryonic Matter (CBM) Experiment at FAIR%
\thanks{Presented at XXIX$^{th}$ International Conference on Ultrarelativistic Nucleus-Nucleus Collisions (Quark Matter 2022), Krakow, Poland, April 4-10, 2022}%
}
\author{Kshitij Agarwal (for the CBM Collaboration)
\address{Physikalisches Institut, Eberhard Karls Universit\"at T\"ubingen, Germany}
}
\maketitle
\begin{abstract}
The CBM experiment at the Facility for Antiproton and Ion Research (FAIR) aims to explore the QCD phase diagram in the region of high net-baryon densities using nucleus-nucleus collisions ($\sqrt{s_{NN}} = 2.9 - 4.9$~ GeV). CBM will be utilizing peak interaction rates of up to 10 MHz and an advanced triggerless data acquisition scheme to provide it with an unique access to rare physics probes required to accomplish CBM's physics goals. This contribution will summarise CBM's progress in terms of its physics performance simulations and the status of the comprising detector sub-systems, including their involvement in FAIR Phase-0 activities.
\end{abstract}
  
\section{Introduction to CBM-FAIR}
\label{Section1}

The Facility for Antiproton and Ion Research (FAIR) \cite{Durante:2019hzd} is a new particle accelerator facility currently under construction which aims to examine cosmic matter in the laboratory. Schwer-Ionen-Synchrotron-100 (SIS-100) is at the heart of FAIR with heavy-ion beams to be injected from the existing SIS-18 ring accelerator at GSI Helmholtz Centre for Heavy Ion Research in Darmstadt. The Compressed Baryonic Matter (CBM) experiment \cite{CBM:2016kpk} is one of the four scientific pillars of FAIR and will utilise beams from SIS-100 (gold beams with kinetic energies between 2-11~AGeV; $\sqrt{s_{NN}} = 2.9 - 4.9$~GeV) to study matter at conditions comparable to astrophysical objects and events \cite{Huth:2021bsp}. Therefore, CBM's physics goals are centered to decipher the properties of Quantum Chromodynamics (QCD) matter at high net-baryon densities and to understand the equation-of-state of dense nuclear matter, the possible phase transition from hadronic to partonic phase, and chiral symmetry restoration. The respective physics observables include sub-threshold strangeness production, collective flow of emitted particles, excitation function of low-mass dileptons and charm production amongst many others. A comprehensive review of CBM's physics goals and observables are in Ref.~\cite{Friman:2011zz}. CBM aims to access these `rare probes' at up to 10~MHz high beam-target interaction rates by using novel self-triggering, free-streaming and radiation-hard readout electronics. These features are complemented by inspection of full data stream in real-time to select and reconstruct events online.

\begin{figure}[b]
\centering
\includegraphics[width=0.95\columnwidth]{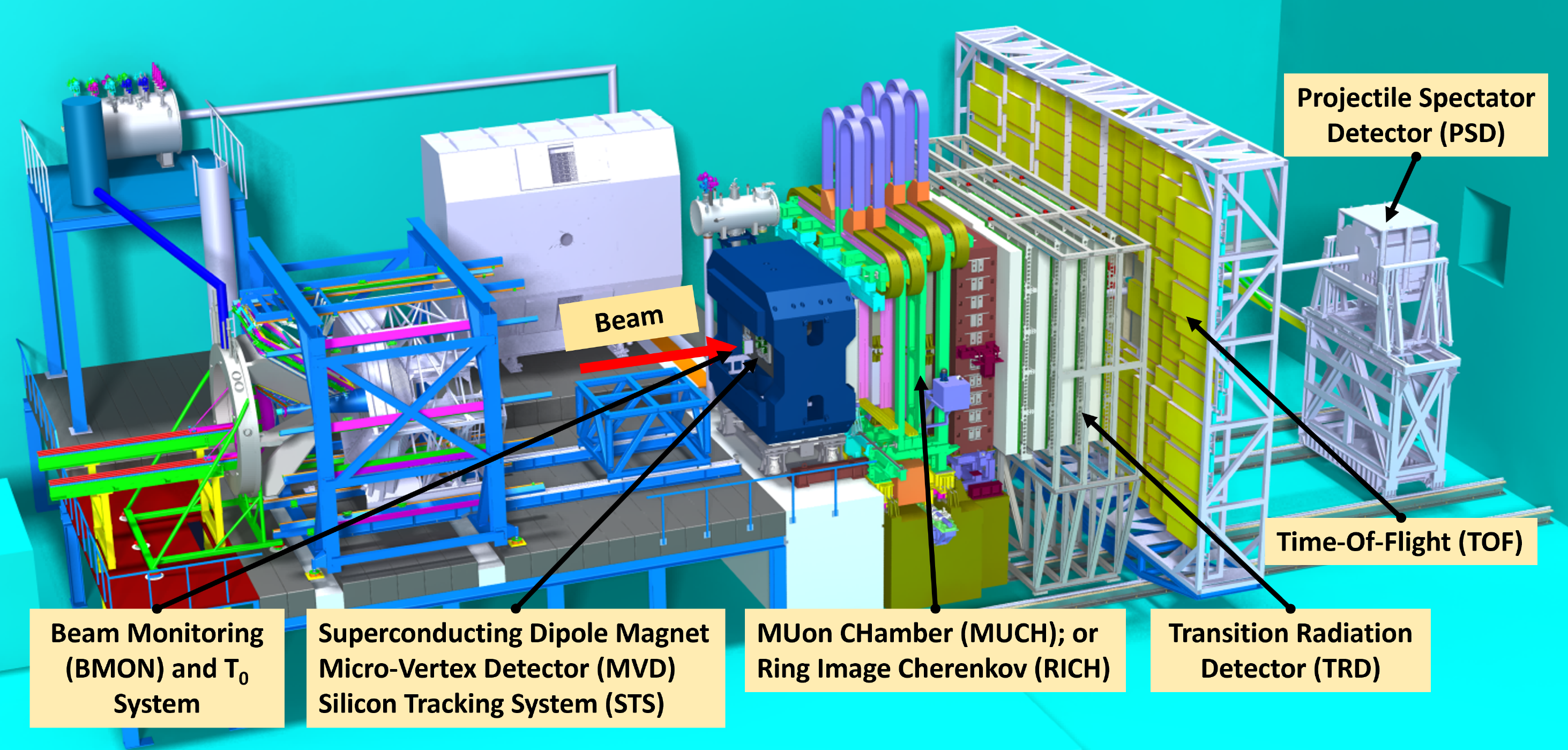}
\caption{CAD view of the CBM (right) and HADES experiment (left). SIS-100 beam will come from left to right of the image (Figure: P. Dahm, GSI Darmstadt).}
\label{fig:figure1}
\end{figure}

CBM is designed as a fixed-target experiment (depicted in Fig.~\ref{fig:figure1}) with an angular acceptance of $2.5^\circ<\theta<25^\circ$ and is capable to measure hadrons and leptons in both their production channels. Located inside the superconducting dipole magnet are the Micro Vertex Detector (MVD) and the Silicon Tracking System (STS) tasked to resolve the secondary vertex of short-lived open-charm mesons and provide momentum determination of charged particles respectively. Dielectron identification is performed by the Ring Imaging Cherenkov (RICH) Detector with the UV photon detector planes based on Multi-Anode Photo Multiplier Tubes. Dimuon identification is done by the Muon Chamber (MuCh) System containing Gas Electron Multipliers and bakelite Resistive Plate Chambers sandwiched between hadron absorber plates. One of CBM's standout features is the interchangeability of RICH and MuCh, allowing CBM to systematically measure dileptons in both production channels for the same detector acceptance. The Transition Radiation Detector (TRD) is also used for pion suppression, particle tracking, and identification of light nuclei and fragments via specific energy loss. Charged hadrons detection is performed by the Time-of-Flight (TOF) wall composed of Multi-Gap Resistive Plate Chambers (MRPCs) located about 7~m downstream. The Projectile Spectator Detector (PSD) used to determine the collision centrality and reaction plane orientation.

\section{Update on Physics Performance Studies}
\label{Section2}

The particle identification (PID) procedures developed for CBM play a central role in benchmarking CBM's physics performance described in subsequent subsections \ref{Section2_1}-\ref{Section2_3}. The Cellular Automaton (CA) method based track finder reconstructs charged particles' trajectories in the STS. This, in combination with the PID information from ToF and TRD, is used to further identify hadrons, light fragments and hypernuclei under the Kalman Filter Particle Finder (KFPF) package. This is further complemented by collision centrality determining methods by using charged particle multiplicities in STS and the energy of spectator fragments to provide a complete picture of the underlying collision.

\subsection{Hadrons and Strangeness}
\label{Section2_1}

\begin{figure}[b]
\centering
\subfigure[]{\label{fig:Vassiliev}\includegraphics[height=0.17\paperheight]{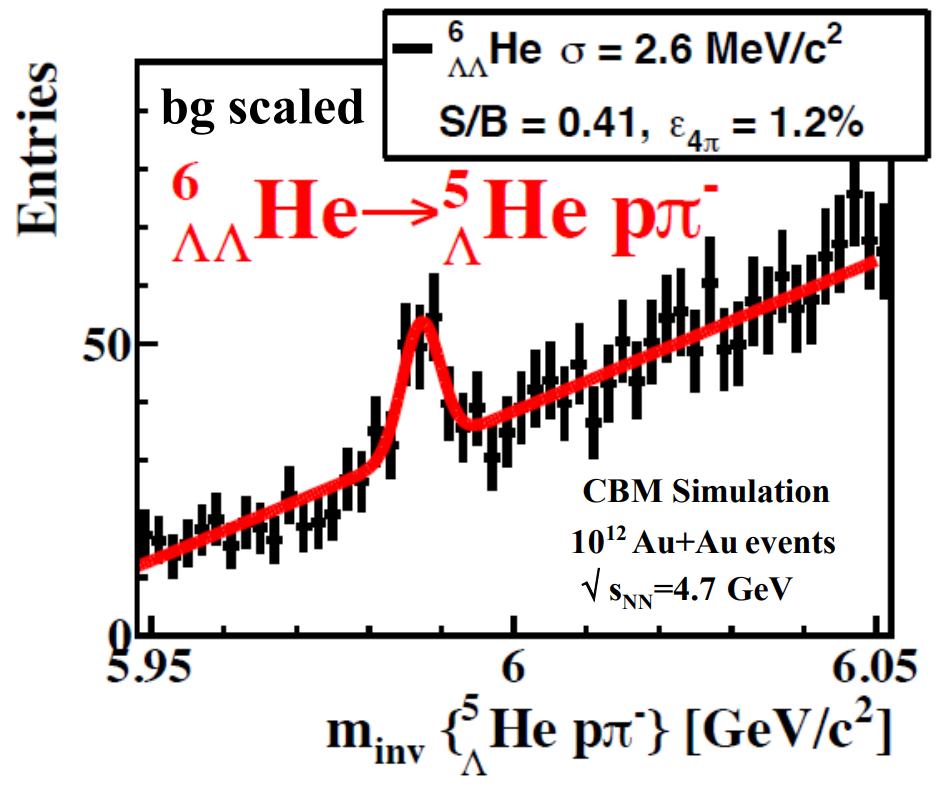}}
\subfigure[]{\label{fig:Khan}\includegraphics[height=0.17\paperheight]{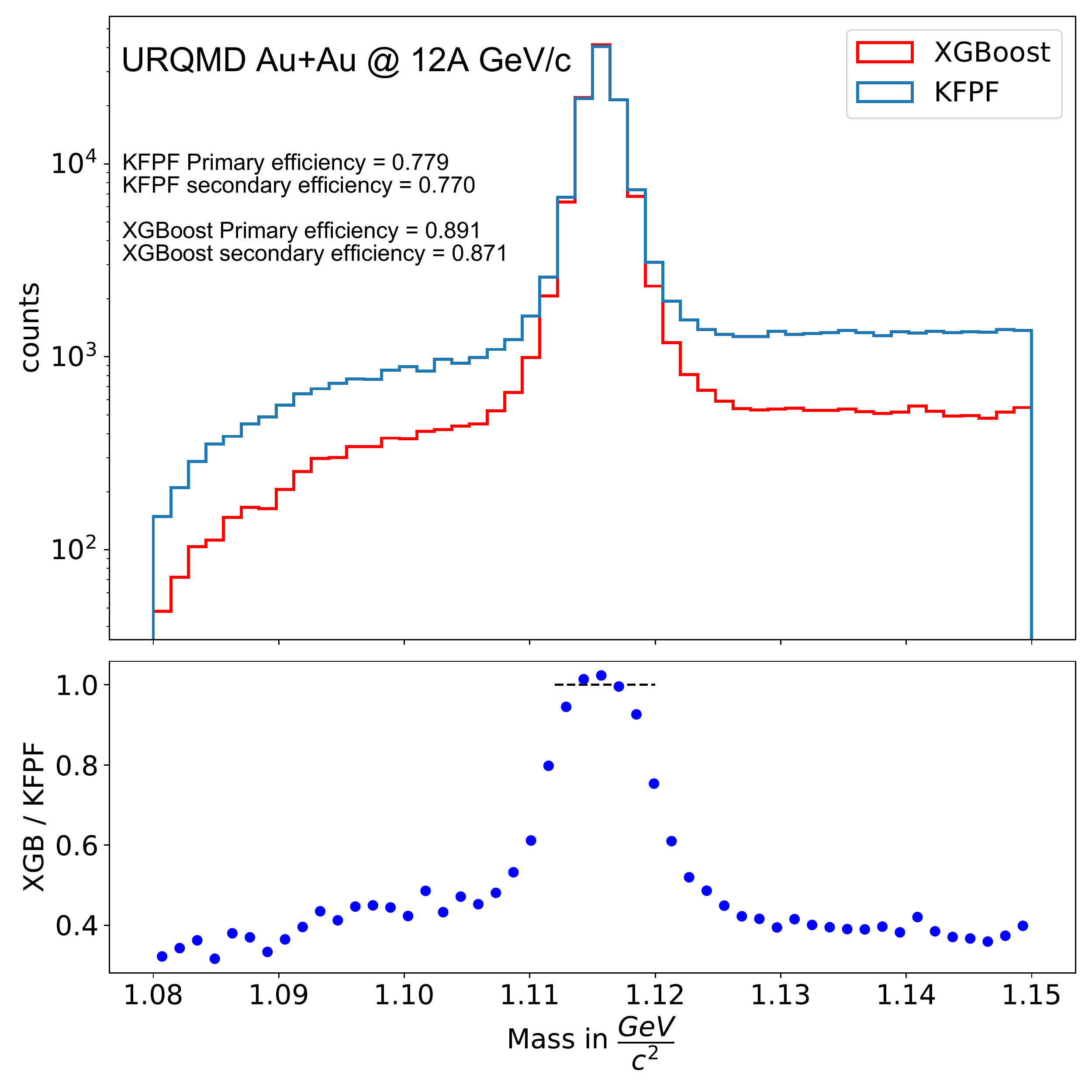}}
\caption{(a) Reconstructed invariant mass distribution of double $\Lambda$ hypernuclei - $^6_{\Lambda\Lambda}He$ (Figure: I. Vassiliev, GSI Darmstadt). (b) Comparison of $\Lambda$ invariant mass distributions for particle selection criteria based on KFPF in blue and XGBoost Machine Learning Model in red (Figure: S. Khan, EKU T\"ubingen).}
\label{fig:figure2}
\end{figure}

Tools for the multi differential physics analysis have been prepared for doing a comprehensive analysis of strange hadrons and hypernuclei (see Fig.~\ref{fig:Vassiliev}). Further optimisation of reconstruction yields of $K_s^0$, $\Lambda$ and $\Xi^-$ have been done by using Machine Learning (ML) techniques to successfully demonstrate high signal purity and efficient rejection of the combinatorial background by $>40\%$ for same efficiency (see Fig.~\ref{fig:Khan}). Additionally, CBM's capabilities to reconstruct proton's transverse momentum spectrum within $5\%$ systematic error for different underlying EoSs in the Three-fluid Hydrodynamics-based Event Simulator (THESEUS) have been shown.

\subsection{Correlation, Flow and Fluctuation}
\label{Section2_2}

Multi-differential analysis of the $v_1$ flow harmonic for $K_s^0$, $\Lambda$ and $\Xi^-$ hyperons have been performed as a function of rapidity and transverse momentum for different centrality classes. This flow analyses is further complemented with feasibility studies using multiparticle azimuthal correlations. This analyses is challenging due to CBM's fixed-target geometry and energy range (non uniform acceptance and low multiplicities), but feasibility in terms of symmetric cumulants (SC($m,n$)) was successfully shown (see Fig.~\ref{fig:Bilandzic}). Feasibility studies to conduct measurements of proton-proton and pion-pion correlations via femtoscopic measurements have also been performed. Further femtoscopic analysis with higher statistics and realistic cuts is ongoing for precise reconstruction of source properties.

\subsection{Dileptons}
\label{Section2_3}

\begin{figure}[t]
\centering
\subfigure[]{\label{fig:Bilandzic}\includegraphics[height=0.15\paperheight]{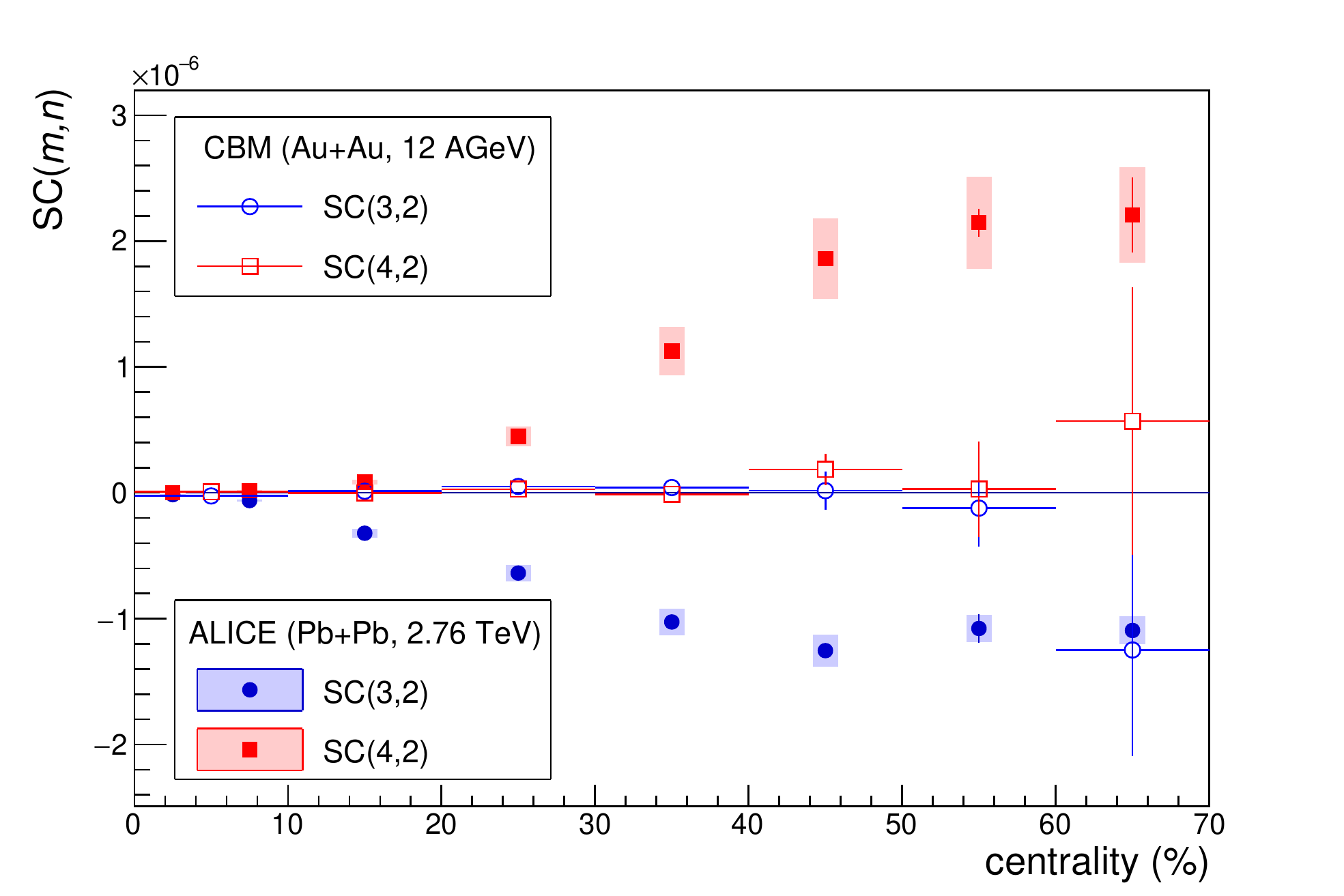}}
\subfigure[]{\label{fig:Chatterjee}\includegraphics[height=0.15\paperheight]{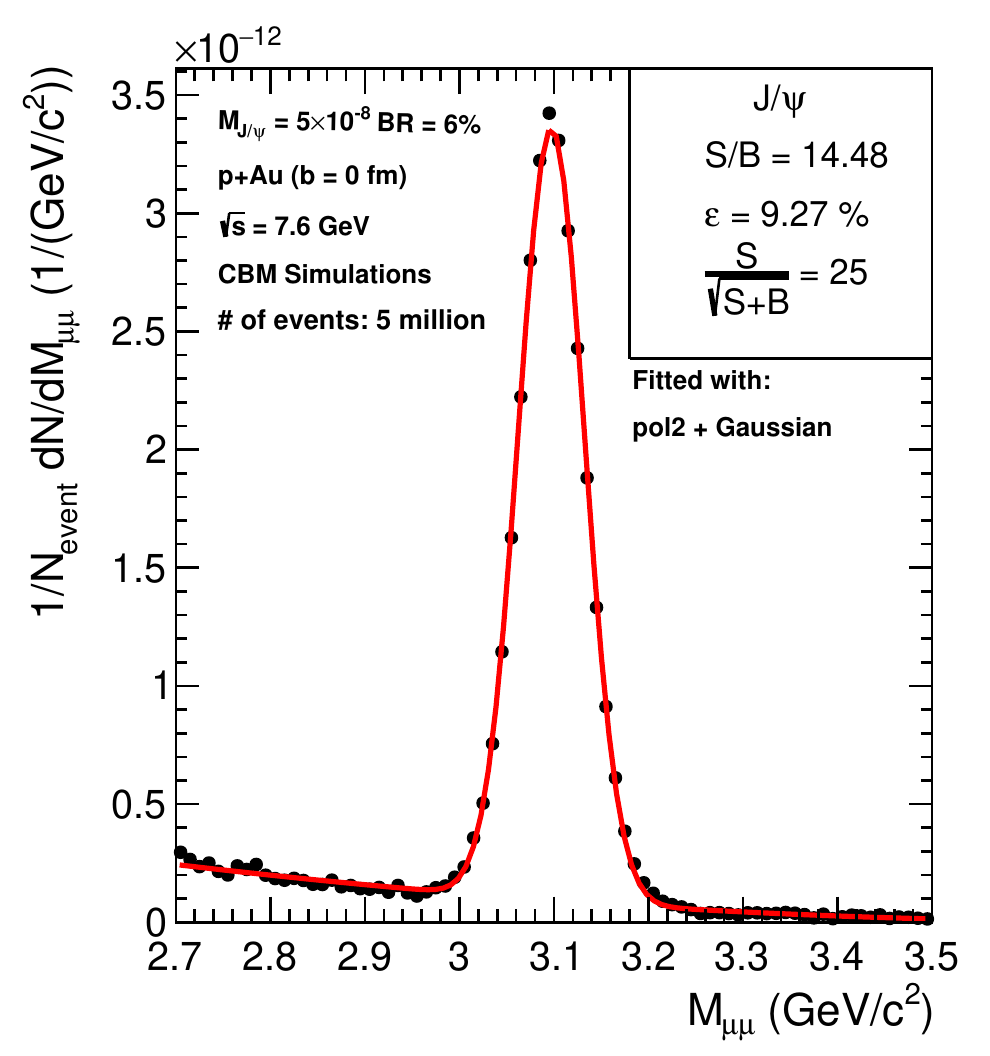}}
\caption{(a) Variation of symmetric cumulants (SC(3,2) and SC(4,2)) with centrality and their comparison with ALICE measurements (A. Bilandzic, TU M\"unchen). (b) Reconstructed $J/\psi$ invariant mass spectra with ($\sigma_M(J/\psi)=35$~MeV/c$^2$) at $\sqrt{s_{NN}} = 7.6$~ GeV p+Au collision (Figure: S. Chatterjee, Bose Institute Kolkata).}
\label{fig:figure3}
\end{figure}

Performance studies, in terms of reconstruction of freezeout cocktail have been done for dileptons (in both production channels $e^+e^-$ and $\mu^+\mu^-$)  with realistic detector geometries, material budget, and response over the entire SIS-100 energy range. This has allowed to access the underlying thermal signal and reconstruct clear peaks for the low mass vector mesons ($\sigma_M(\omega)=14$~MeV/c$^2$) with a pair detection probability of $\approx10\%$. A particular emphasis has also been placed on using sub-threshold charmonium measurements to study cold nuclear matter effects p+A collisions and to study pQCD inspired models at low energies (see Fig.~\ref{fig:Chatterjee}).

\section{Detector Updates and FAIR Phase-0}
\label{Section3}

Almost all CBM detector sub-systems have completed their R\&D phases, with their Technical Design Reports available in Ref.~\cite{CBMweb}. MVD has shown production readiness for the pre-production sensor version. The engineering and production readiness of various STS components was reviewed for pre-production to start in late-2022. MuCh has built and tested real size GEM and RPC chambers, and have conceptually reviewed their mechanical aspects. RICH has accomplished its ``first of series test" for their electronics and is constructing a full-size photo-camera demonstrator. TRD has submitted the proposal for their inner high-rate inner modules with novel 2D readout pads. TOF has undertaken an extensive campaign to determine their rate limits and mitigation efforts towards observed aging are ongoing. Technical designs of the target mechanism, beam pipe and beam monitors have been developed. Evaluation of the impact on the project of sanctions imposed on Russia, especially regarding the superconducting dipole magnet, the PSD and the CBM's timeline is presently ongoing. 

\begin{figure}[h]
\centering
\subfigure[]{\label{fig:mCBM1}\includegraphics[height=0.11\paperheight]{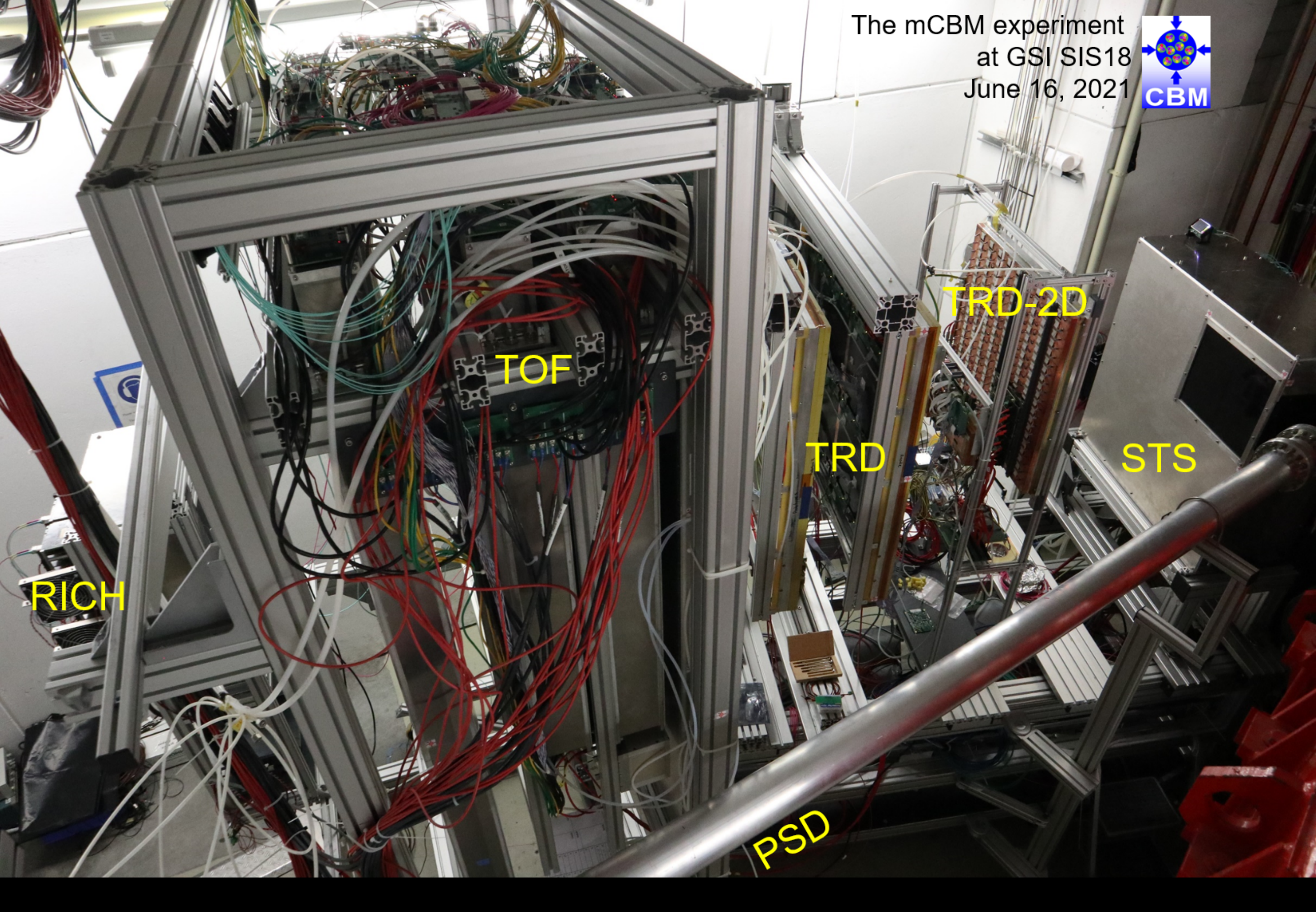}}
\subfigure[]{\label{fig:mCBM2}\includegraphics[height=0.11\paperheight]{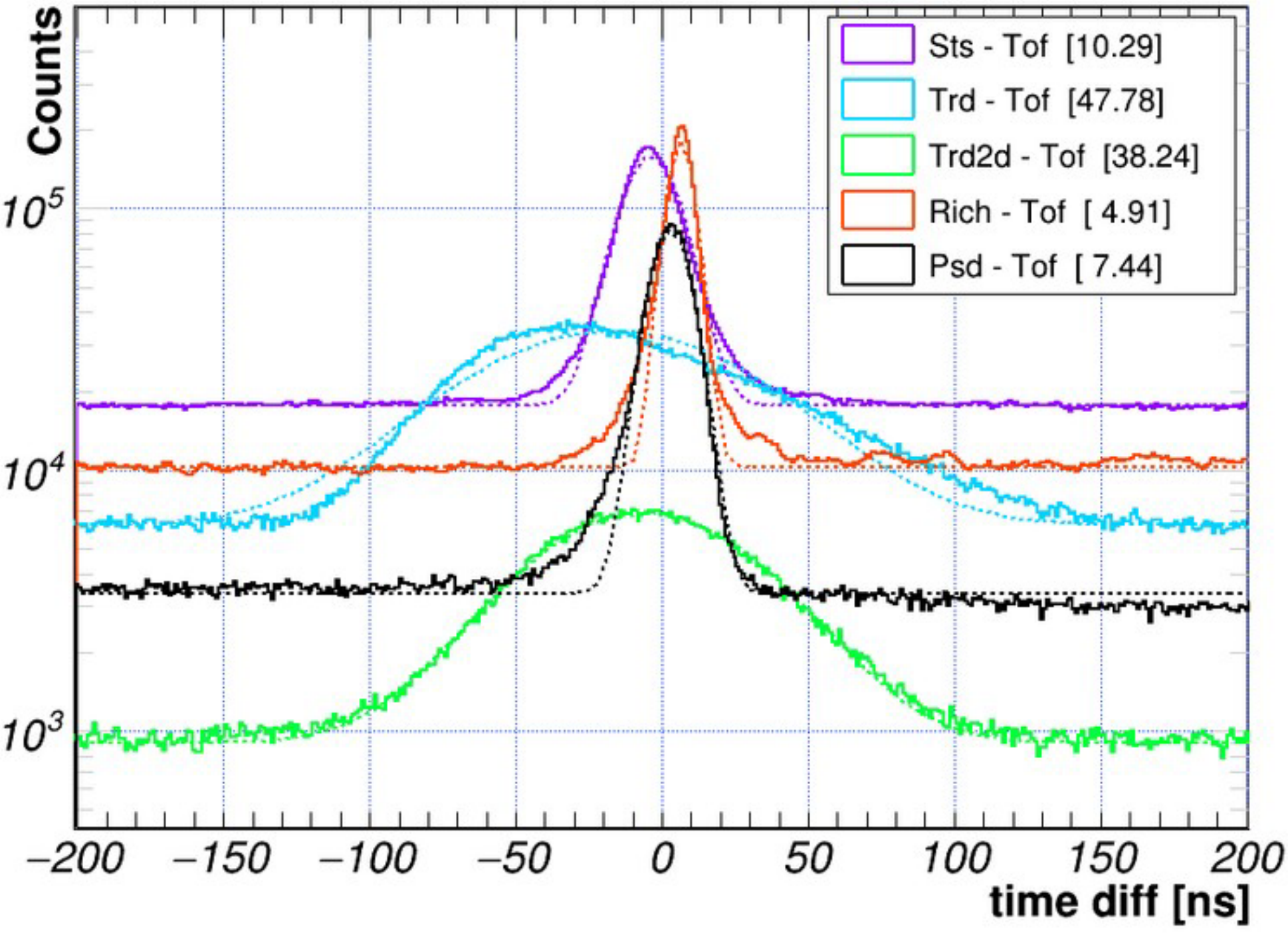}}
\caption{(a) mCBM setup for the 2021 campaign (Figure: C. Sturm, GSI Darmstadt). (b) Time synchronicity between all sub-systems for O+Ni collisions at 2.0 AGeV at about 1 MHz collision rate (Figure: A. Weber, JLU Giessen).}
\label{fig:figure4}
\end{figure}

Different aspects of CBM have been and are continued to be thoroughly tested in FAIR's precursor programme - FAIR Phase 0. The 2021 campaign of mCBM$@$SIS-18 at GSI \cite{mCBM:220072} has successfully demonstrated the triggerless-streaming readout and data transport of CBM with a setup composed of prototype components of all CBM's detector sub-systems with the readout chain interfaced with the Common Readout Interface (CRI) boards (see Fig.~\ref{fig:figure4}). Moreover, detailed high-rate tests have been performed for radiation tolerance validation of various sub-system's components. Other FAIR Phase-0 activities include: (i) the successful implementation of CBM's KFPF package in STAR's express analysis chain and endcap Time-of-Flight (eTOF) comprising 36 CBM-TOF modules for BES-II campaign \cite{Kisel:2020lpa,Deppner:2020vom}; (ii) HADES-RICH upgrade with 428 MAPMTs from CBM-RICH \cite{HADES:2020CBMPR}.

\section{Conclusion and Outlook}
\label{Section4}

CBM, with its foreseen capabilities can map out the high-density region of the QCD phase diagram by studying conditions similar to the core of neutron stars and can help usher a new era of multi-messenger physics. This is substantiated by the recent physics performance studies focusing on the rare probes required to achieve this goal. All detector sub-subsystems are progressing well towards the strategic objective to be ready for series production. The active participation in FAIR Phase-0 has contributed substantially to CBM's understanding of the underlying sub-systems. mCBM after verifying CBM's readout concept is currently undertaking benchmark runs targeting $\Lambda$ baryon reconstruction in Au+Au collisions at 1.24 AGeV and Ni+Ni collisions at 1.93 AGeV.

\section{Acknowledgements and Funding}
\label{Section5}
The CBM experiment is being developed by a collaboration consisting more than 400 persons from 47 institutions and 11 countries. CBM is supported by national funding agencies of CBM member institutions. The author acknowledges the support from the German Federal Ministry of Education and Research (Project-ID 05P19VTFC1) and the Helmholtz Graduate School for Hadron and Ion Research (HGS-HIRe).

\bibliographystyle{unsrt}
\bibliography{refs}

\end{document}